\documentclass[letterpaper]{article}
\usepackage[pdftex]{graphicx}
\usepackage{lineno}
\usepackage{amsmath}
\usepackage{hyperref}

\usepackage[left=0.98in, right=0.98in, top=1.0in, bottom=1.0in]{geometry}

\newcommand\beq{\begin{equation}}
\newcommand\eeq{\end{equation}}

\newcommand\bmat{\begin{bmatrix}}
\newcommand\emat{\end{bmatrix}}

\newcommand{\Sel}{\textit{Selector~}}
\newcommand{\Seq}{\textit{Sequence~}}
\newcommand{\Par}{\textit{Parallel~}}
\newcommand{\Suc}{\textit{Success~}}
\newcommand{\Fail}{\textit{Failure~}}

\newcommand{\Ca}{Ca$^{++}$ }
\newcommand{\UW}{University of Washington}

\begin{document}
\setpagewiselinenumbers        %  Line numbers for edits to drafts.
\modulolinenumbers[1]          %  number every N lines

% \linenumbers                   %  start numbering lines here

\title{Behavior Trees as a Representation for Medical Procedures}

\author{Blake Hannaford$^1$ \and Randall Bly$^2$ \and Ian Humphreys$^2$ \and Mark Whipple$^2$}
\date{
$^1$ Biorobotics Laboratory, Department of Electrical Engineering, \UW \\
$^2$ Department of Otolaryngology / Head And Neck Surgery, \UW \\[0.2in]
\today}

\maketitle
\begin{center}
Submitted to Journal of the American Medical Informatics Association, Aug 2018
\end{center}

\begin{abstract}
{\bf Objective} Effective collaboration between machines and clinicians requires flexible data structures to represent medical processes and
clinical practice guidelines.  Such a data structure could enable effective turn-taking between human and automated components of a
complex treatment, accurate on-line monitoring of clinical treatments (for example to detect medical errors), or
automated treatment systems (such as future medical robots) whose overall treatment plan is understandable and auditable
by human experts.
 
 {\bf Materials and Methods} 
Behavior trees (BTs) emerged from video game development as a graphical language for modeling
intelligent agent behavior.  BTs have several properties which are attractive for modeling
medical procedures including human-readability, authoring tools, and composability. 

 {\bf Results} This paper
will illustrate construction of BTs for exemplary  medical procedures and clinical protocols
\footnote{We are pleased to acknowledge support from National Science Foundation grant \#IIS-1637444 and collaborations on that
project with Johns Hopkins University and Worcester Polytechnic Institute.}.

{\bf Discussion and Conclusion}
Behavior Trees thus form a useful, and human authorable/readable bridge between clinical practice guidelines and AI systems.

\end{abstract}

%%%%** Section 1
\section{Background and Significance}

A major trend in medicine is the development of
evidence-based \href{https://academic.oup.com/intqhc/article/28/1/122/2363781}{clinical practice guidelines}\cite{kredo2016guide} and protcols  which standardize and
communicate best practices in treatment or management of a patient's condition.
Increasingly, the events and data arising from such treatment are being captured in electronic medical records,
logged by medical devices during use, or collected by cameras,  sensors, and ubiquitous computing devices (e.g. cell phones).
Machine learning is good at identification of relationships between variables in large  data sets,
but lacks a means of determining causality.

%  key intro material from Mark Whipple
Over the past two decades a number of computer-interpretable guideline (CIG) formalisms have been developed\cite{OpenClinical}, including rule-based (Arden syntax), logic-based (PROforma), network-based (PRODIGY), and workflow-based (GUIDE) models.  These allow medical knowledge, in the form of narrative clinical guidelines or informal flowcharts, to be transformed into a model that can interpretable by health information technologies and applied to individual patient data for clinical decision support\cite{MorPeligGuideline2006}.  While the specifics vary, these models typically represent a guideline as a nested plan or task, consisting of primitives such as decisions, actions, patient states, and execution states\cite{wang2001representation} and often including a flowchart. For example, the Guideline Interchange Format (GLIF3) is an object-oriented specification which includes a flowchart type task network consisting of different types of steps, such as decision, patient-state, branch, synchronization and action, as well as a well developed data and query model\cite{boxwala2004glif3}. In the PROforma model, clinical guidelines are represented as plans, which contain tasks such as actions, decisions and enquiries\cite{vollebregt1999study}.  The GuideLine Acquisition, Representation and Execution (GLARE) model is a graph based structure in which query, work, decision, and conclusion actions are defined as nodes\cite{anselma2006towards,terenziani2008applying}.

All of these systems feature comprehensive data descriptions but pay a corresponding price in terms of complexity.   Although AI techniques have been used, for example in semantic checking of timing and periodicity\cite{terenziani2008applying}, less attention has been paid to application of these descriptions to automated reasoning, sensor information processing, or robotics.

In a panel discussion paper on the state of AI in Medicine\cite{Patel2009Coming}, Peter Szolovitz writes
\begin{quotation}
``I think it is a great
challenge to build better modeling tools that permit
the integration of human expertise (recognizing its
fallibility) with machine learning methods that
exploit a huge variety of available data."
\end{quotation}
and
\begin{quotation}``Human expertise, developed over centuries
of experience and experimentation, cannot be
discarded in the hope that it will all be re-discovered
(more accurately) by analyzing data."
\end{quotation}

\paragraph{Objective}
In this paper we apply Behavior Trees (BTs) to the representation of existing medical procedures, developed by the profession, but at multiple levels of abstraction.  The goal is to
integrate advanced medical procedural knowledge with AI systems, via
stimulating exploration of BTs in medical informatics systems.

The purpose of this paper is to illustrate the process of converting previously published and validated clinical procedures and protocols to BTs.  There are  differing opinions among experts about
clinical protocols.  Thus, the focus of this paper is on the process of converting to a BT and
not on which specific clinical protocol to use.  Diverse examples were chosen for this paper to
illustrate the applicability of BTs to both medical procedures and medical management protocols.

\paragraph{Background}
Prior to about 2010, the term “Behavior Tree” was used idiosyncratically by several authors, but around that time a body of literature began to emerge around a tree model of behaviors used by the video game industry for AI-based non-player characters\cite{halo,lim2010evolving}.  These BTs assume that units of intelligent behavior (such as decisions or units of action) can be described such that they perform a piece of an overall task/behavior, and that they can determine and return a 1-bit result indicating success or failure.  These units are the leaves of BTs.
The level of abstraction of BT leaves is not specified by the BT formalism and varies from one application to another or within a single
BT.
In the context of
medicine, BT leaves could be diagnostic or theraputic steps such as the administration of a blood test or a small
step of a surgical procedure such as tying a knot.   In describing patient management, that might occur over several days, a BT leaf
might describe a sub-procedure such as to perform a biopsy, but that biopsy could in-turn be broken down into its
own BT.

In medical robotics, researchers are turning attention to augmentation of the purely teleoperated surgery of existing
systems such as the daVinci$^{TM}$ surgical robotic system (Intuitive Surgical, Sunnyvale, CA) with intelligent
functions\cite{BRL278,BRL279}.  In this context, BT leaves could represent such functions  as a guarded move,
a precision cutting action, acquisition of an ultrasound image, creation of a plan, etc.
Earlier medical robotics systems containing automation, such as Robodoc\cite{kazanzides1992architecture}
addressed represented task sequences with
scripting languages.

Recent literature has applied BTs to UAV control\cite{ogren2012increasing}, humanoid robotic control\cite{tumova2014maximally}, and human-robot cooperation in manufacturing\cite{paxton2017costar}.
Theoretical classification of BTs has been conducted by several authors\cite{TRO17Colledanchise} which  formally relates BTs to Finite State Machines (FSMs). BTs have advantages of modularity and scalability with respect to finite state machines.  Other theoretical studies have related BTs to Hybrid Dynamical Systems\cite{colledanchise2014behavior}, humanoid robotic behavior\cite{tumova2014maximally}, and have developed
means to guarantee correctness of BTs\cite{colledanchise2017synthesis}.
Software packages and Robotic Operating System (ROS, \cite{quigley2009ros})
implementations\footnote{\url{https://github.com/miccol/ROS-Behavior-Tree}} are now available\cite{marzinotto2014towards}.
Several of the above references
have ample introductory material and examples of BT concepts.

\paragraph{Materials and Methods}

When representing a process with BTs,
the analyst breaks the task down into modules which are the  leaves of the BT.
Each BT node must return either \Suc or \Fail when called by its parent node.
All higher level nodes in the BT
define composition rules to combine the leaves including: \Seq, \Sel, and \Par node types.
A \Seq node defines the order of execution of leaves and returns \Suc if all leaves succeed in order,
returning \Fail at the first child failure.
A \Sel node (also called ``Priority" node by some authors) tries leaf behaviors in a fixed order,
returns success  when a node succeeds,  and returns failure if all leaves fail.
A \Par node starts all its child nodes concurrently and returns success if a specified fraction of its
children return success.
Further refining the behavior tree,
{\it Decorator} nodes have a single child and can modify behavior of subsequent branches
with rules such as ``repeat until $X>0$".

Applied to robotics,
BTs have been explored in the context of humanoid
robot control
\cite{marzinotto2014towards,colledanchise2014performance,bagnell2012integrated} and recently as a modeling language for intelligent robotic surgical procedures \cite{hu2015semi}.

Many clinical practice guidelines or algorithms are today represented as flowcharts, using a familiar notation
similar to the \href{https://www.iso.org/standard/11955.html}{ISO 5807:1985} standard for computer program flowcharts.
One use of such a diagram of a medical procedure is as a real-time reference to guide practice
(for example an emergency airway algorithm
posted on the wall of an ER).
We do not propose the use of BTs for this purpose. At this time it appears that a clincian would need some
additional training to read a BT properly.
In this paper we instead introduce  BTs as a means to represent medical procedures (or clinical practice guidelines)
in a form that is both machine readable, useful with theoretical constructs such as
hybrid system theory, as well as human authorable and readable.
A data structure with these properties could enable effective turn-taking between human and automated components of a
complex treatment, accurate on-line monitoring of clinical treatments (for example to detect medical errors), or
automated treatment systems (such as future medical robots) whose overall treatment plan is understandable and auditable
by human experts.
In contrast, automated medical procedures achieved by machine learning (e.g. \cite{van2010superhuman}) cannot be understood by even expert humans.

% \newpage
%%%%** Section 2
\section{Results}
\subsection{Example 1. Blood Draw}

%%%%** Figure 1
\begin{figure}\centering
\includegraphics[width=1.0in]{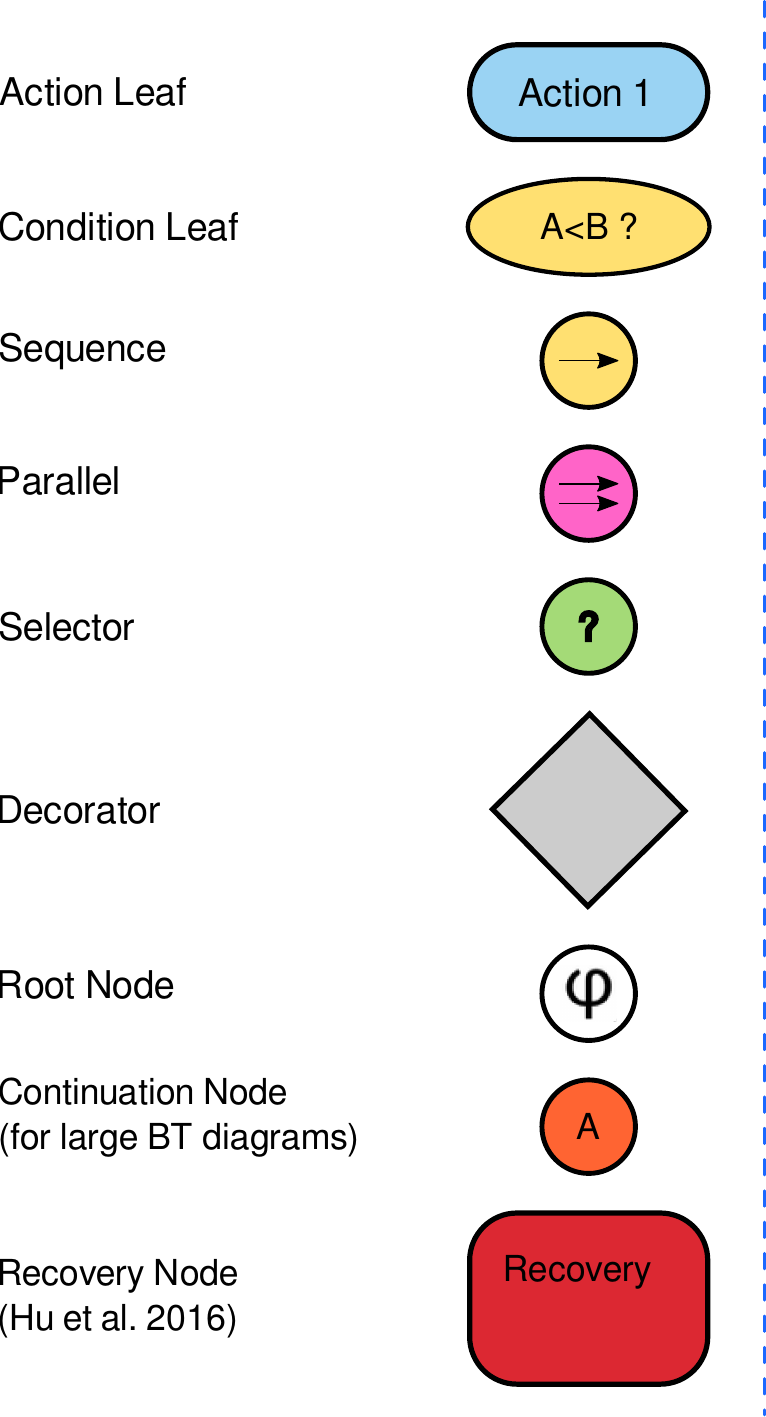} \hspace{0.35in}
\includegraphics[width=4.750in]{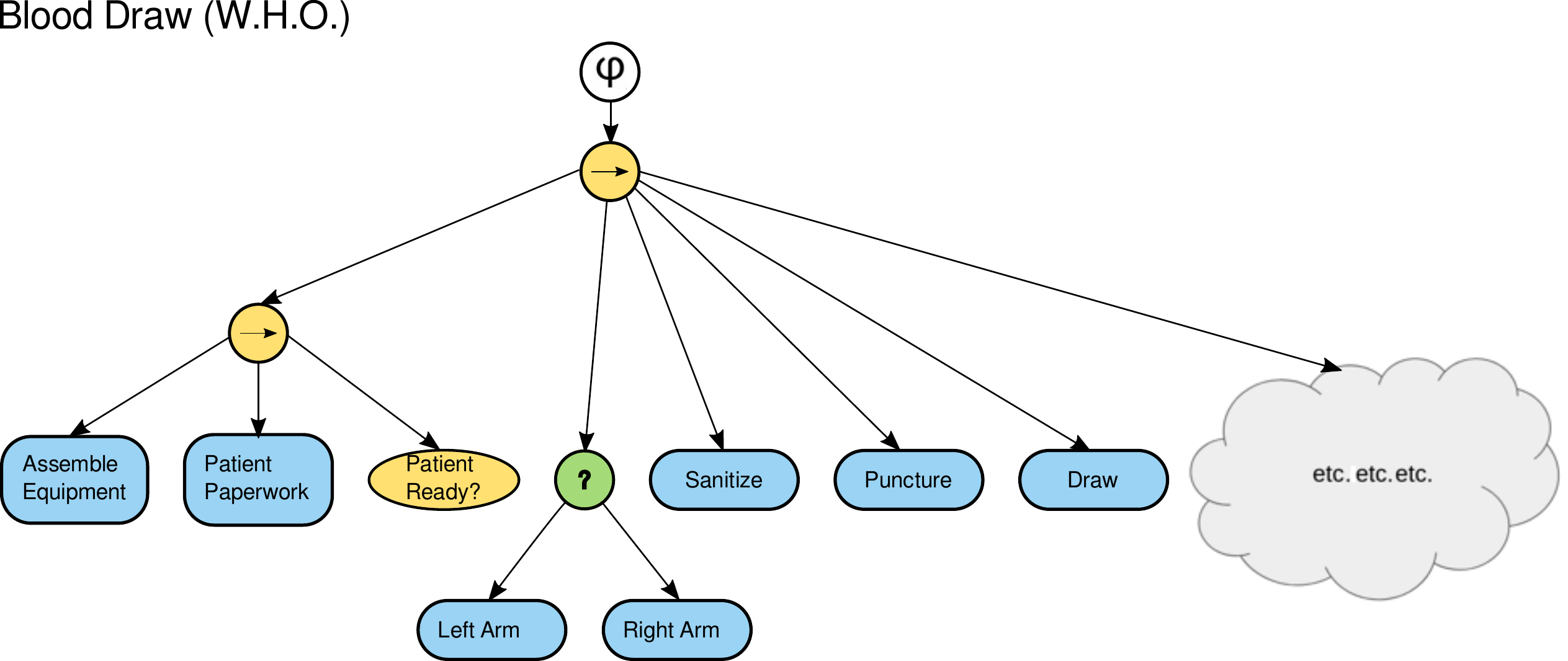}
\caption{Behavior Tree constructed (using BT symbols at left) for the basic blood draw procedure documented by the World Health Organization\cite{WHO_bestpractices}.}\label{BloodDrawBTfig}
\end{figure}

The World Health Organization issues a best practices document\footnote{\url{http://who.int/infection-prevention/tools/injections/drawing_blood_best/en/}}
on drawing blood for medical tests (phlebotomy)\cite{WHO_bestpractices}.
This over-100-page document gives many details for each step of what is
mostly a serial process with few branches.  A BT representing
the first several steps of this process was developed and is represented
in Figure \ref{BloodDrawBTfig}.  It is worth re-emphasizing that we make no statement about the
effectiveness or appropriateness of the medical procedures described in the examples selected here,
but rather we use them to illustrate the BT notation.  Nor do we intend to criticize the appropriateness of
the cited flowcharts for their intended purposes.

The root (top) node of a BT, $\Phi$, encapsulates task start, task end, and the overall task \Suc or \Fail status.
For this procedure, its only child
is a \Seq node (\includegraphics[width=0.15in]{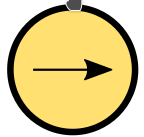}) indicating that execution will be passed to
each child in sequence from left to right as shown, with \Fail returned
by the node if any child returns \Fail (the BT root always has one child).
The first child of the  main \Seq node is also
a sequence node which secures equipment and paperwork, and assesses
the overall readiness of the patient.
In this and subsequent diagrams, leaves of the tree are actions (blue), or queries
(yellow) to indicate a logical test or sensing operation. Query nodes return
\Suc or \Fail   if a condition is satisfied (in this case if the patient is ready).
Blue leaves of the tree indicate tasks that are
physically performed.   The second child of the main \Seq node
is a \Sel node (\includegraphics[width=0.15in]{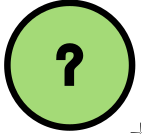}) in which the phlebotomist determines whether
or not a suitable vein is present in the left or right arm.   If neither
arm shows a suitable vein then the \Sel node will fail and
that failure will propagate up to the Sequence and in turn to the tree itself.
This tree could be simplified with equivalent meaning if the lower left \Seq node
is deleted and its three children connected to the top-level \Seq node.

%%%%** Figure 2
\begin{figure}[h]\centering
\includegraphics[width=0.75\textwidth]{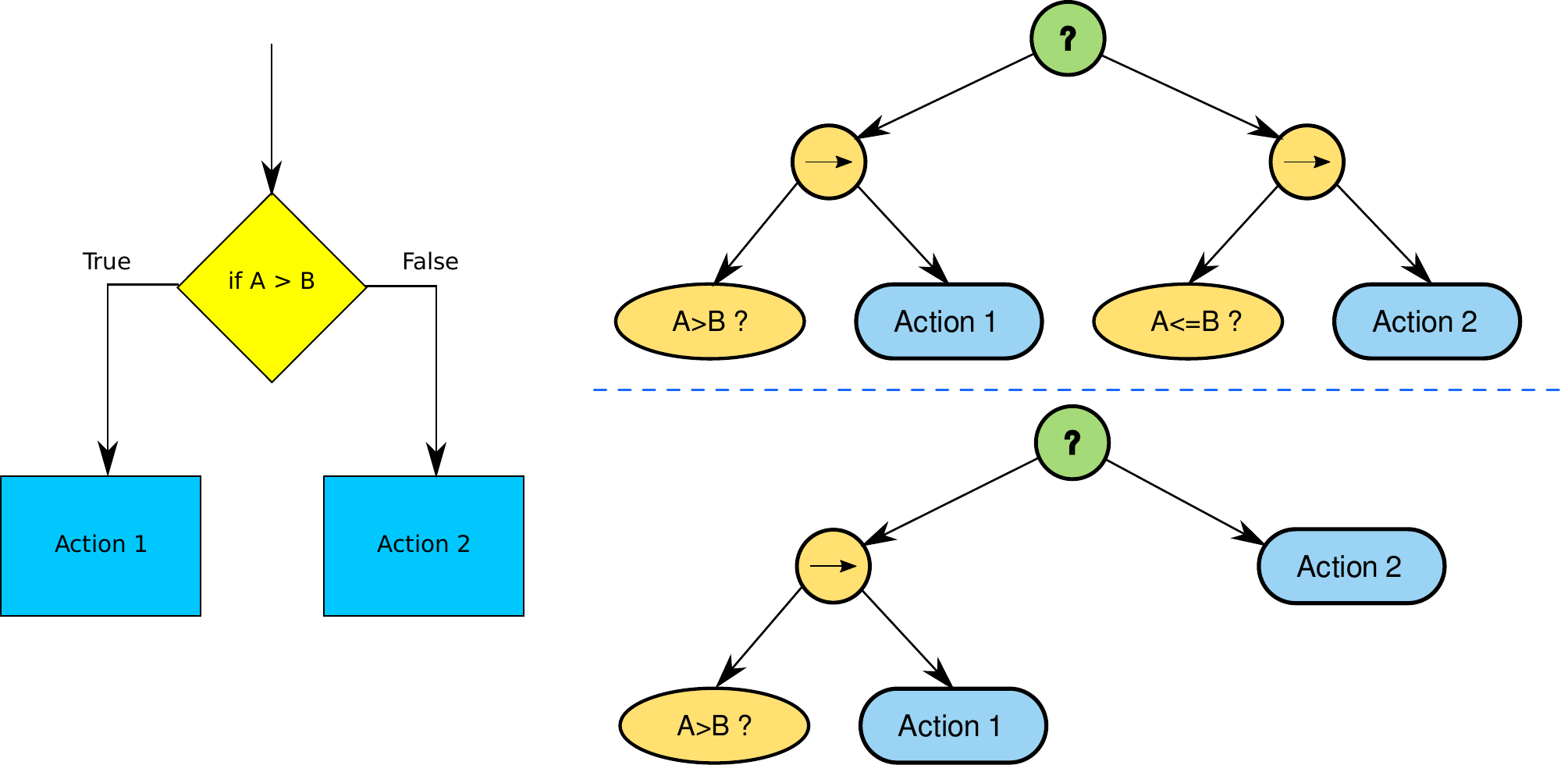}
\caption{Mapping of a flow chart if-then-else condition (left) to two  equivalent BTs (right, separated by dashed line). Child nodes (the Condition and the Actions) could be replaced with arbitrarily detailed sub-BTs.}\label{IfThenElseMapping}
\end{figure}

%%%%** Section 3
\subsubsection{Process of BT Creation}
So far it is a manual process to convert an existing published algorithm into a BT.
When an algorithm is published as a flow-chart, this process is straightforward when aided by
a few conventions.

Sequences in traditional flowcharts are traditionally represented by a vertical chain of blocks.
Each procedural block may be mapped to a BT leaf (although flowcharts do not require a \Suc or \Fail
decision from each block).  These leaves are then connected by a \Seq node (\includegraphics[width=0.15in]{sequence_BTsym.png}).

For branching, many flowcharts contain a diamond shape to indicate
``{\it if-then-else}''.  To straightforwardly convert if-then-else blocks into a BT,
replace the block with a \Sel node (\includegraphics[width=0.15in]{selector_BTsym.png}).
The first child of the \Sel is a \Seq
in which the first child of that \Seq is a test for the condition. The test returns \Suc
if the condition is true and \Fail if the condition is false.
The second child  node represents the
flow chart action(s) for the cases where the condition succeeds.
If the condition fails, control passes to the \Sel node
which then executes a child representing the condition-fail branch.  This mapping is illustrated
in Figure \ref{IfThenElseMapping}.

Graphical user interfaces have been developed for BT authoring, or the BT can be exported from
existing tools like GLARE\cite{terenziani2008applying}.

% \newpage

%%%%** Figure 3
\begin{figure}[t]\centering
\includegraphics[width=0.65\textwidth]{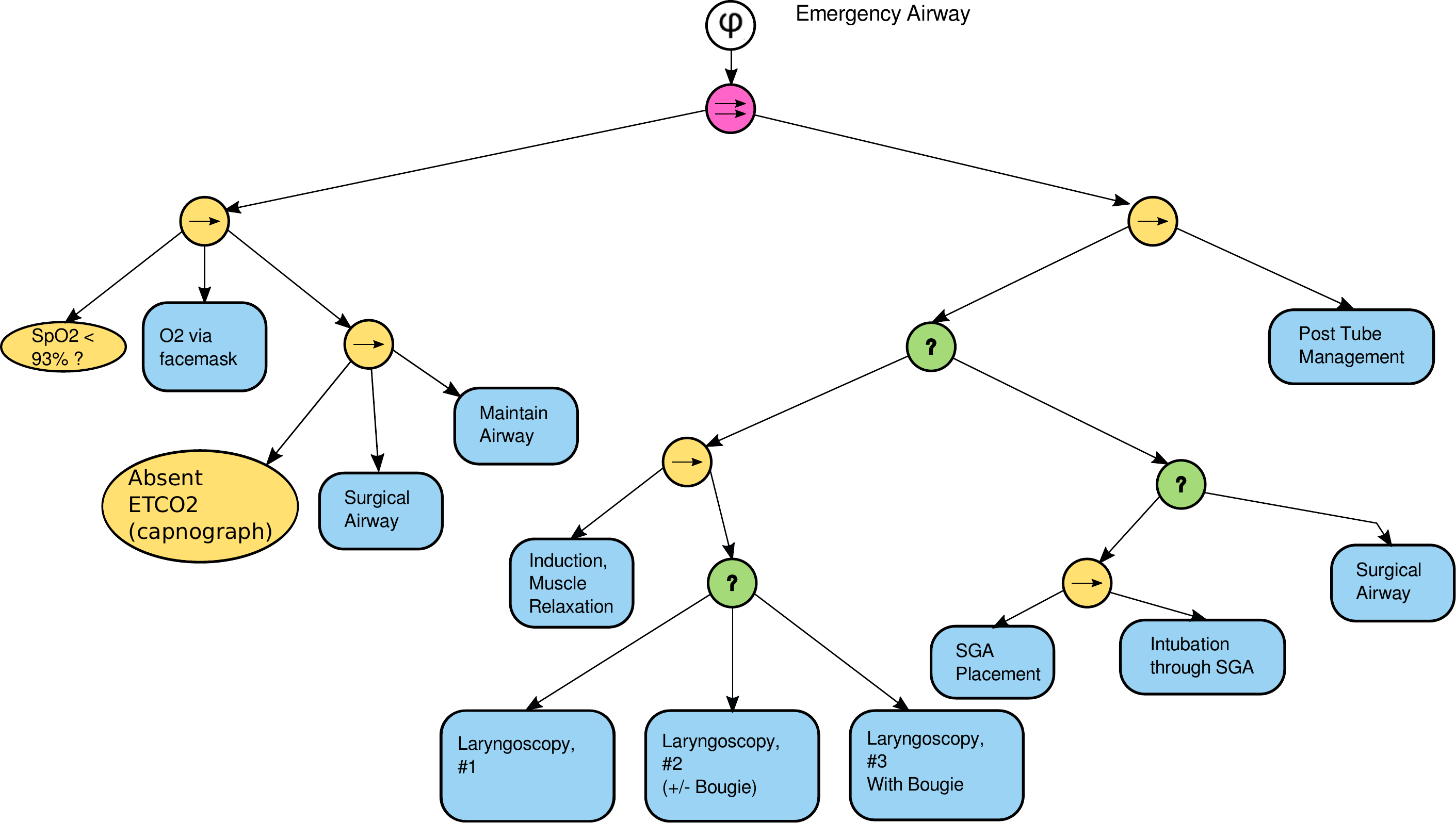}
\caption{BT constructed for the emergency airway procedure of \cite{stephens2009success}.}\label{airwayBTfig}
\end{figure}
%
% %%%%** Figure 4
% \begin{figure}[h]\centering
% \includegraphics[width=0.5\textwidth]{AirwayFlowchart_Davisetal2007.png}
% \caption{Emergency airway establishment procedure, Davis et al. \cite{davis2007effectiveness} (used with permission). }\label{airwayDavisFig}
% \end{figure}

%%%%** Figure 4
\begin{figure}[h]\centering
\includegraphics[width=0.5\textwidth]{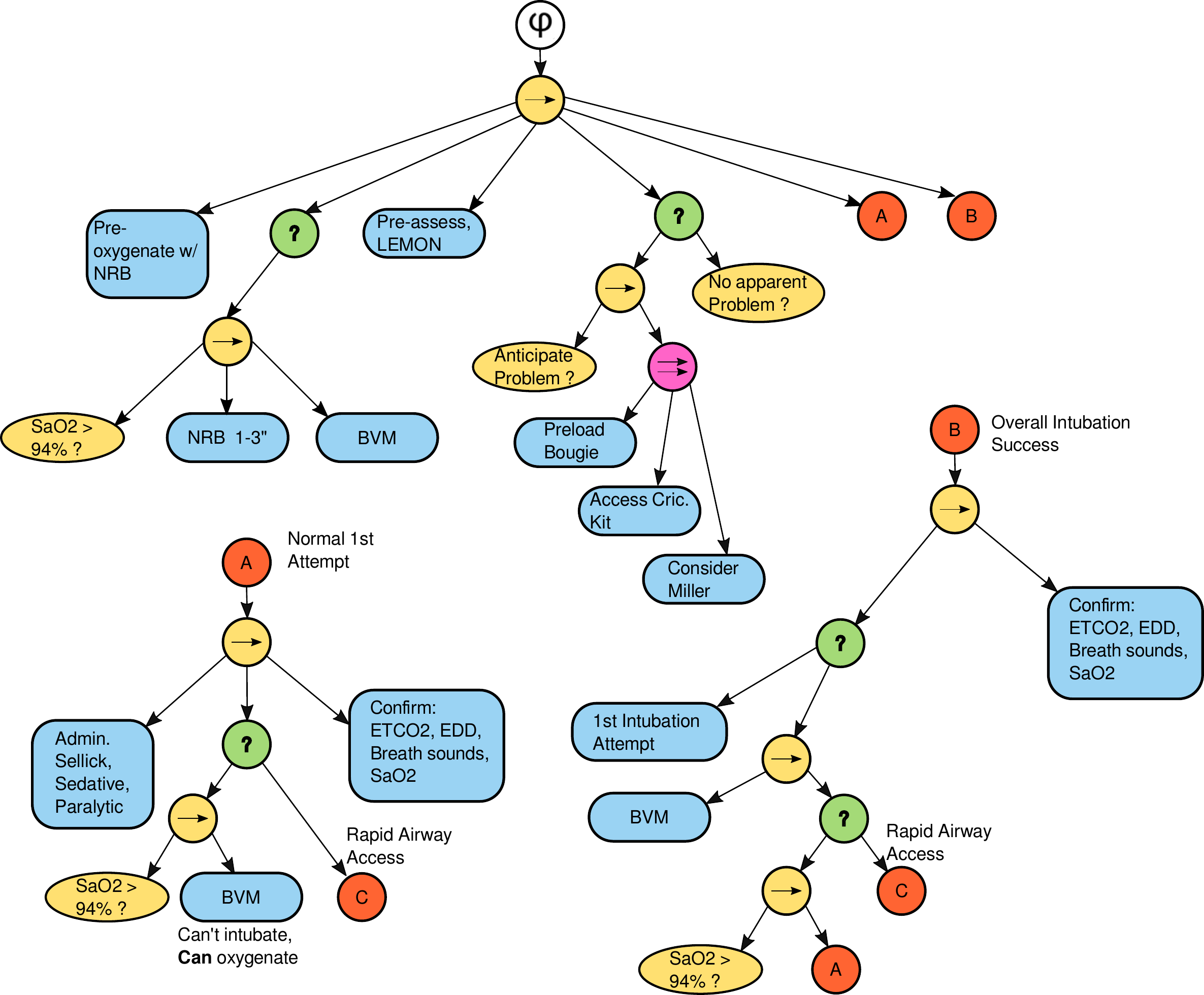}
\includegraphics[width=0.4\textwidth]{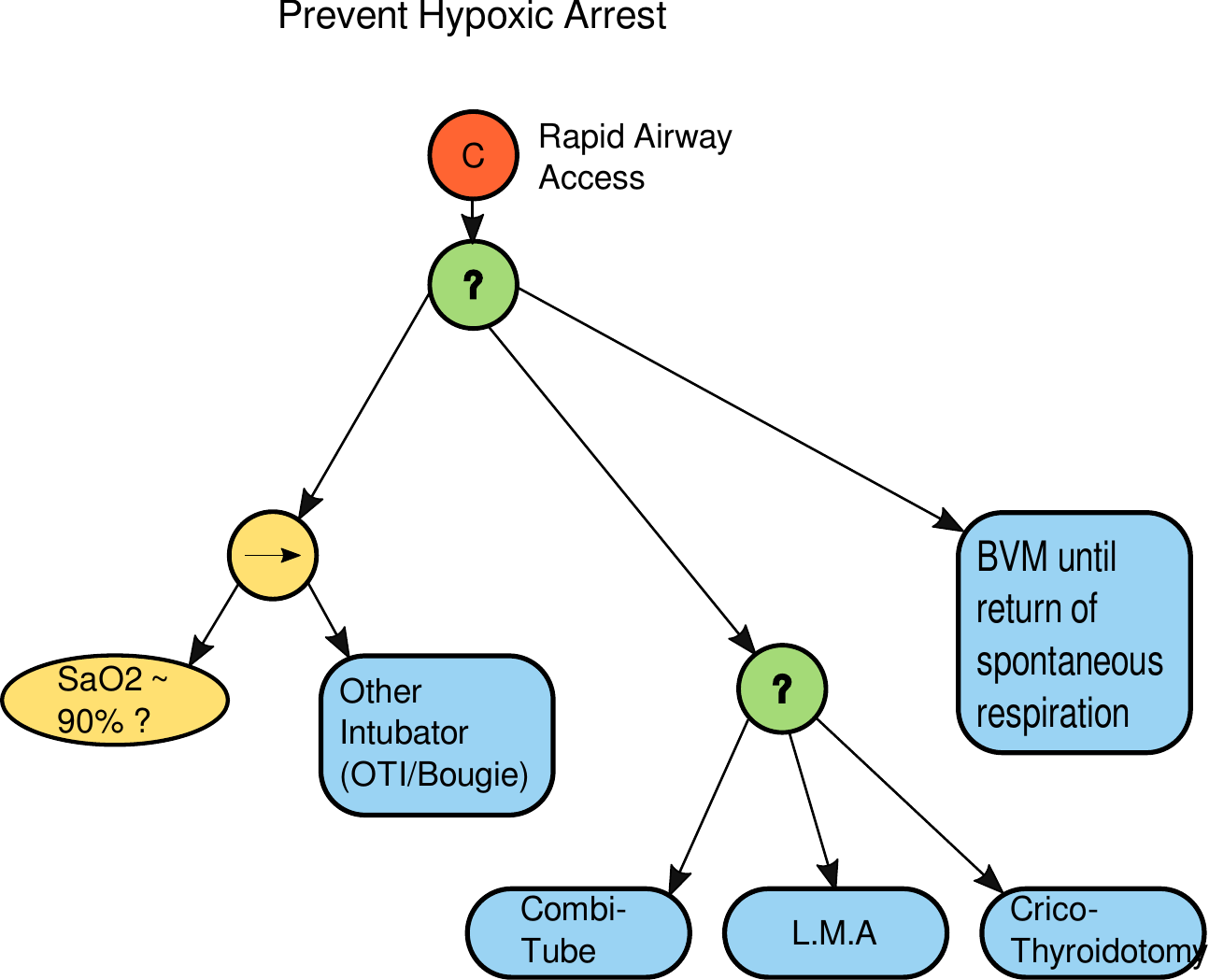}
\caption{BT for emergency airway establishment procedure of Davis et al.\cite{davis2007effectiveness} }\label{airwayDavisBT1Fig}
\end{figure}

%%%%** Section 4
\section{Example 2. Emergency Airway Ventilation}

%
% \begin{figure}[h]
% %\includegraphics[width=3.0in]{ASA_AirwayAlgDecisionTreea.png}
% % \includegraphics[width=3.0in]{Davis_etal_2009_airwaya.png}
% \vspace{1.0in}
% \href{http://www.sciencedirect.com/science/article/pii/S0952818004000480}{[Left]}
% \href{http://www.tandfonline.com/doi/full/10.1080/10903120601023370}{[RIGHT]}
% (Figures pending Copyright permissions)
% \vspace{1.0in}
% \caption{Existing representations of airway establishment include Left: American Society of Anesthesiology\cite{rosenblatt2004airway},
% Right: Davis et al. \cite{davis2007effectiveness} }\label{airwaytradfigs}
% \end{figure}
%

Human life will expire in minutes if the upper airway is blocked.
A medical team thus must quickly follow
a best practice sequence of interventions until airflow is reestablished.
Restoration of airway consists of a rapid succession of increasingly
invasive steps, starting with insertion of a laryngoscope, and, as a last
resort, surgical opening of the airway through crychothyroidotomoy.
The literature on airway restoration algorithms contains many diagrammatic languages for
representation of the airway algorithm.   In one example\cite{stephens2009success}
the flow chart includes an exception in the form of a separate box to the side of the
flowchart containing:

\begin{quotation}
        ``If SpO2 drops to 93\% at any point:
        Facemask + OPA or SGA. If no ETCO2 with best attempts,
        progress to surgical airway."\cite{stephens2009success}
\end{quotation}

This box can is explicitly outside the flowchart but indicates a concurrent
monitoring and interrupt task which is hard to represent in the original selected notation.

We constructed a BT for the airway procedure based on \cite{stephens2009success} and interpreted by
\url{https://emcrit.org/racc/shock-trauma-center-failed-airway-algorithm/}(Figure \ref{airwayBTfig}).
 The first logic node (directly below $\Phi$) is a
\Par node, which indicates that its children should execute concurrently.  The first child of the
\Par node represents the concurrent
monitoring procedure represented as a side box in \cite{stephens2009success}.
The right branch, defining the main algorithm,  contains a sequence node ($\to$).  Its left-most child in
turn is a \Sel node which allows for alternative methods, returning when the first of its children
succeeds. It can be verified that in the normal procedure, the surgical airway procedure
is a last-resort which only is attempted when laryngoscopy (up to 3 attempts)
and intubating Laryngeal Mask Airway (LMA) placement (two attempts) both fail.  LMA and Supraglottic Airway (SGA) are similar equipment.

Compared to the flowchart of \cite{stephens2009success}, the BT is a uniform representation which clearly labels alternative strategies and fallbacks (via the \Sel nodes), and is amenable to direct software
execution (assuming code modules (such as for example ROS nodes) are available for each
leaf.

A second airway establishment procedure was published by Davis et al.\cite{davis2007effectiveness}
using a flow-chart like diagram.
% (Figure \ref{airwayDavisFig}).
A BT for their procedure is given in Figure \ref{airwayDavisBT1Fig}

\clearpage
% \newpage
%%%%** Section 5
\subsection{Example 3. Simulated Tumor Margin Ablation}

%%%%** Figure 5
\begin{figure}[t]\centering
\includegraphics[width=.65\textwidth]{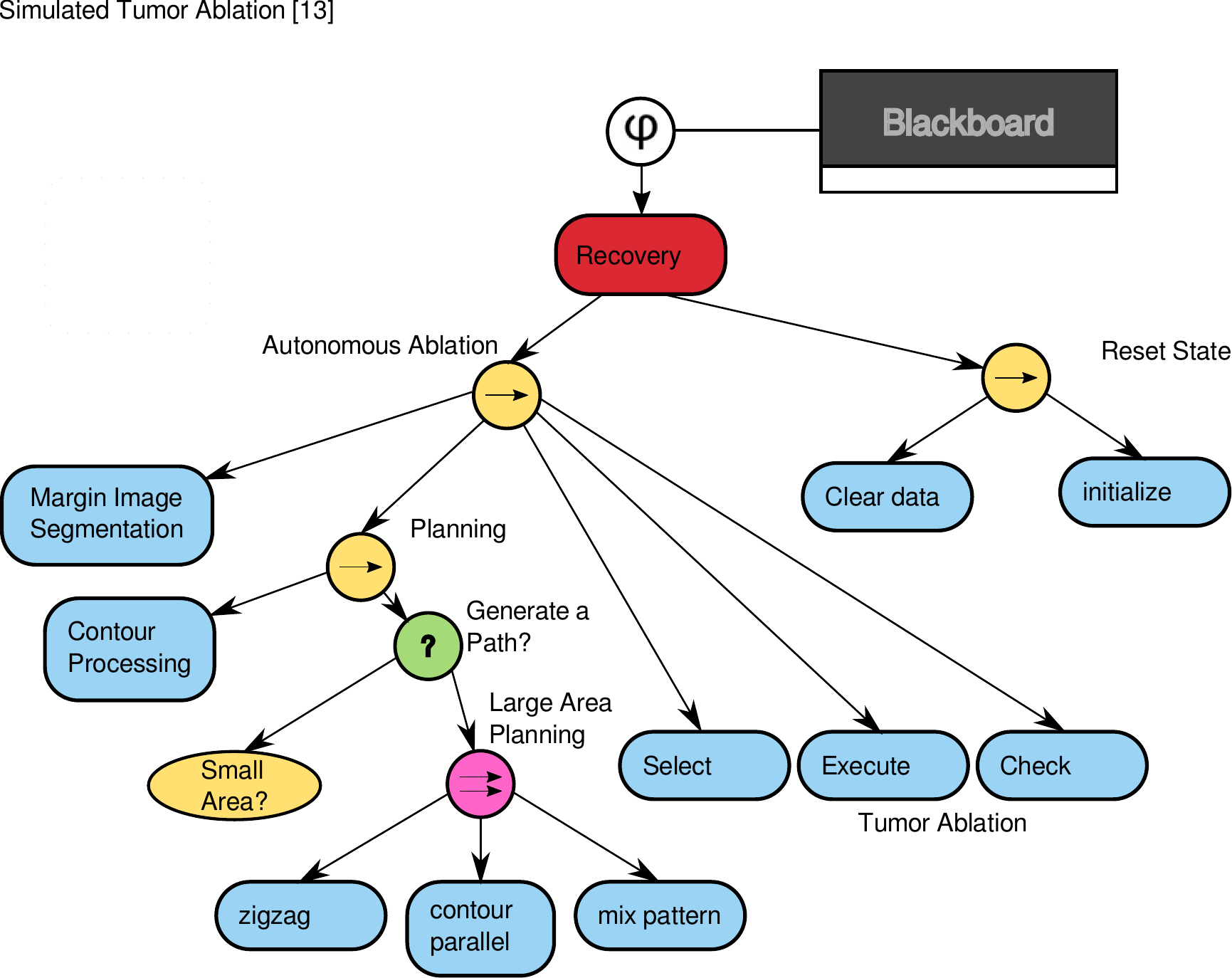}
\caption{BT implemented in software controller for robotic detection and ablation/treatment of positive
tumor margins (redrawn from \cite{hu2015semi}, with permission).
A blackboard data store is commonly
used with BTs to allow leaves to share information.}\label{HuTumorAblationBT}
\end{figure}

In recent bench-top surgical robotics experiments\cite{hu2015semi,hu2017semi}
a system was developed which illustrated a future surgical scenario for
treatment of glioma.  In this scenario, a surgeon will expose the tumor
and manually remove it, but the problem remains of detecting and treating
any remaining tumor material at the edge of the resulting cavity.
In many cancer surgeries, a margin of up to one centimeter is taken around
the tumor to increase the odds that no residual cells are left behind.

In this work Hu et al. assumed the existence of a currently-under-development
biomarker for brain tumors\cite{veiseh2007tumor} which would allow
residual tumor material to be detected through fluorescence.  They
developed a robotic system which could scan the cavity for stimulated fluorescence,
detect a response, and plan and execute one or more treatment plans.

The BT we developed (Figure \ref{HuTumorAblationBT}) performs this task,
and checks up to four planning algorithms (lower left leaves) for appropriateness depending on the area and
shape of the detected fluorescent region.  Notably Hu et al., developed a new type of node, the ``Recovery"
node, which is able to fall back to a recovery tree in the event of a task failure.

Another notable feature of this Medical BT is the action leaf labeled ``Select".
In this
implementation, selecting of the plan from among several computed plans,
was performed by manual input from a surgeon.   Thus the BT framework can
easily incorporate manual steps into a complex and composable procedure.
Furthermore, should an automated function be developed with sufficient
confidence, it can easily be dropped in to the select leaf node of the BT.

% \newpage
%%%%** Section 6
\subsection{Example 4: BT for post operative patient management}

One attractive feature of BTs is that they are equally useful at different levels
of abstraction and time scales.  In this section we describe a BT for post
operative management of Calcium in thyroidectomy patients.   After
thyroidectomy, it is critical that clinicians properly verify that Calcium
regulation by the parathyroid glands is not disrupted by the thyroidectomy.
Patel et al.\cite{patel2018clinical}  describe (using a traditional flow-chart)
a newly validated clinical pathway for postoperative calcium management after
thyroidectomy in pediatric patients.  We have converted this algorithm into a BT
containing 47 leaves.

Besides expanding the applicability of our proposed
theoretical approach to patient management and higher level actions, this
application will challenge our process with a significantly larger BT.

%%%%** Figure 6
\begin{figure}[h]\centering
\includegraphics[width=0.6\textwidth]{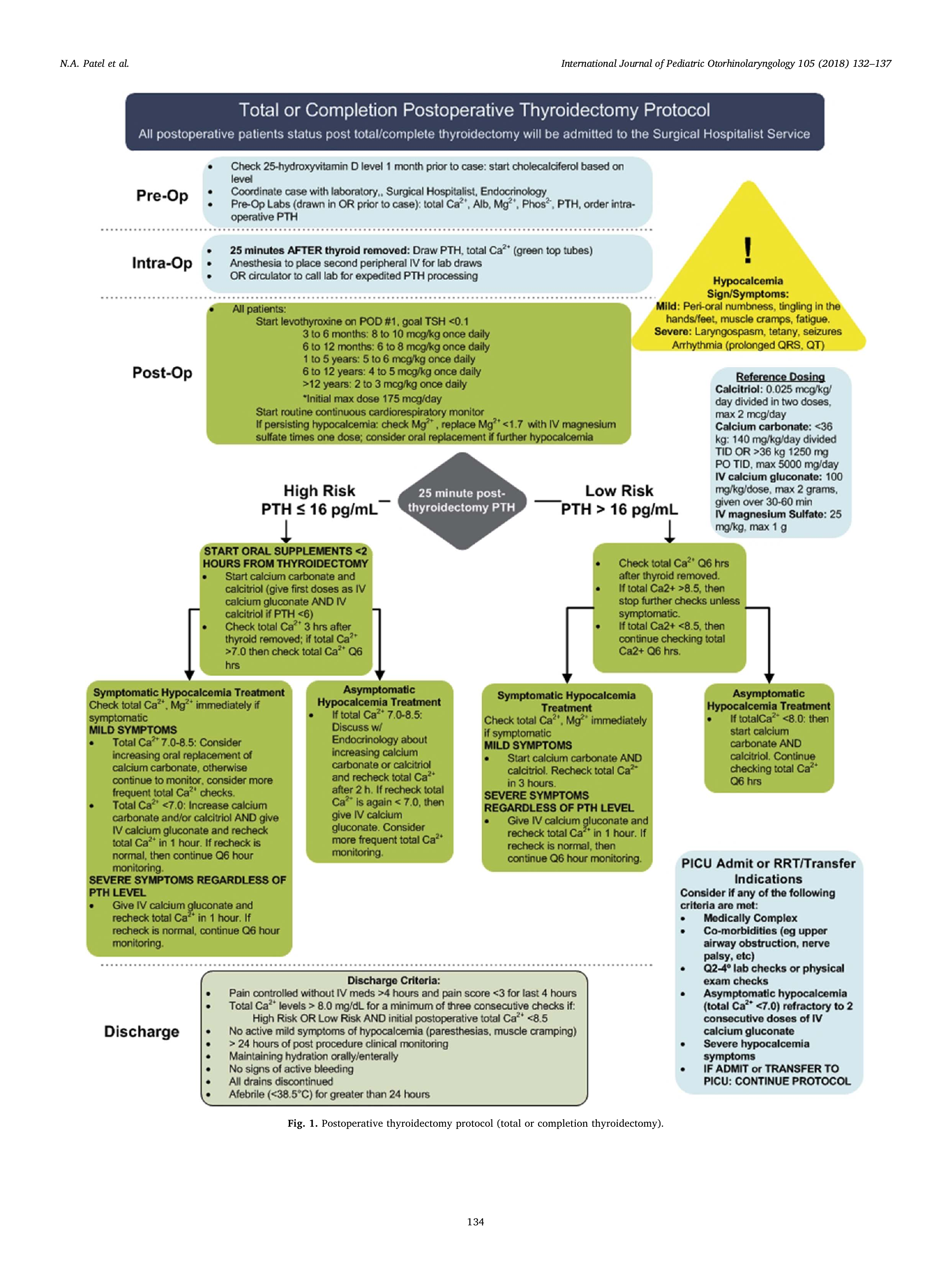}
\caption{Conventional flow-chart (Patel et al.\cite{patel2018clinical}, used with permission)
 post thyroidectomy patient management of Calcium.}\label{PatelFlowchart}
\end{figure}

\begin{figure}[h]\centering
\includegraphics[width=\textwidth]{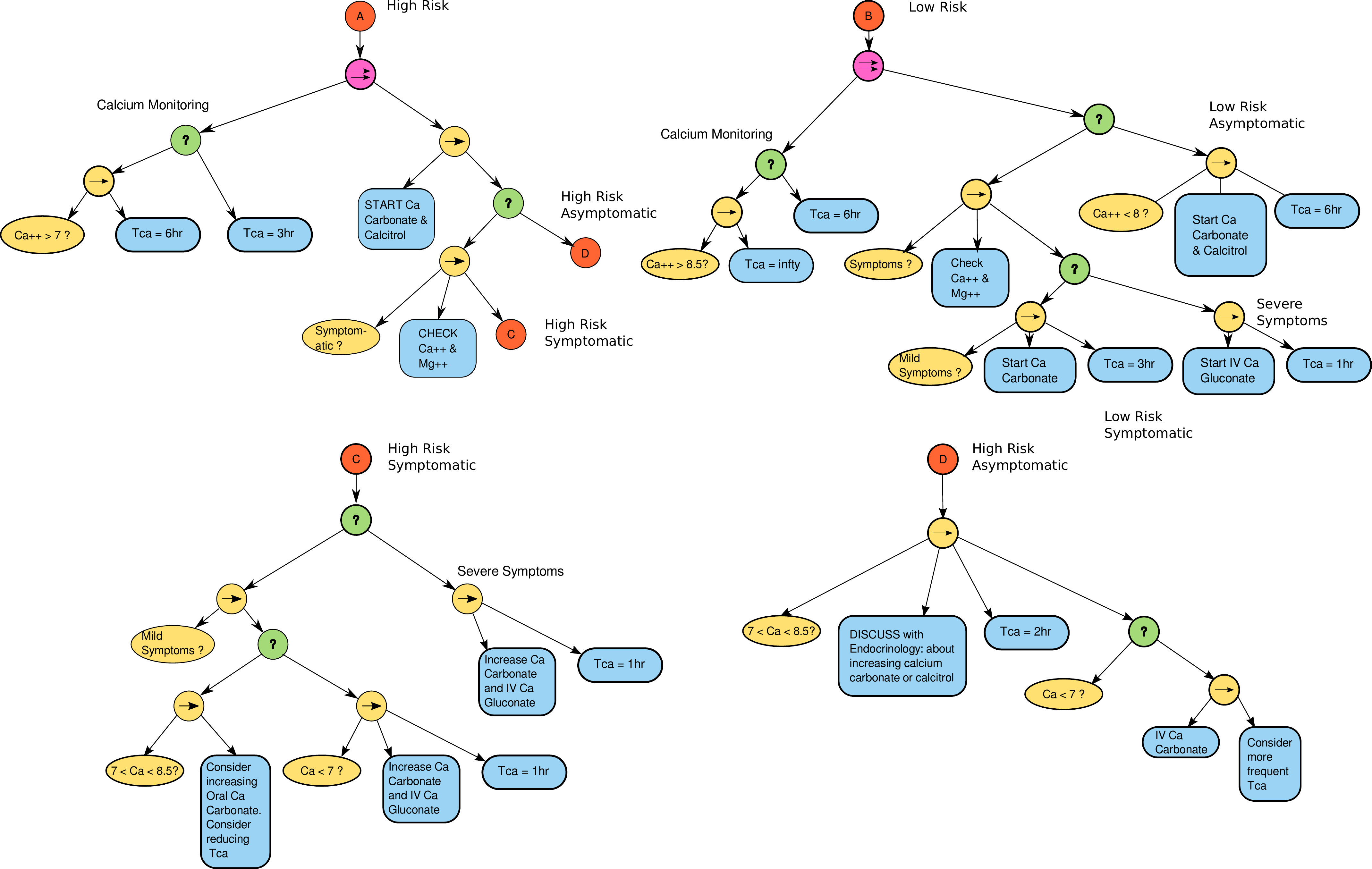}
%  \begin{tabular}{cc}
%  \includegraphics[width=0.5\textwidth]{HypoCalc_A.pdf} &
%  \includegraphics[width=0.50\textwidth]{HypoCalc_B.pdf}  \\
%  \includegraphics[width=0.4\textwidth]{HypoCalc_C.pdf} &
%  \includegraphics[width=0.45\textwidth]{HypoCalc_D.pdf}
%  \end{tabular}
\caption{BTs for variations in  hypo-calcemia treatment in post-operative managment of total  thyroidectomy cases for:
(A) High Risk,
(B) Low-Risk,
(C) High Risk Symptomatic,
(D) High Risk Asymptomatic.}\label{CaBT}
\end{figure}

\begin{figure}[h]\centering
\includegraphics[width=0.7\textwidth]{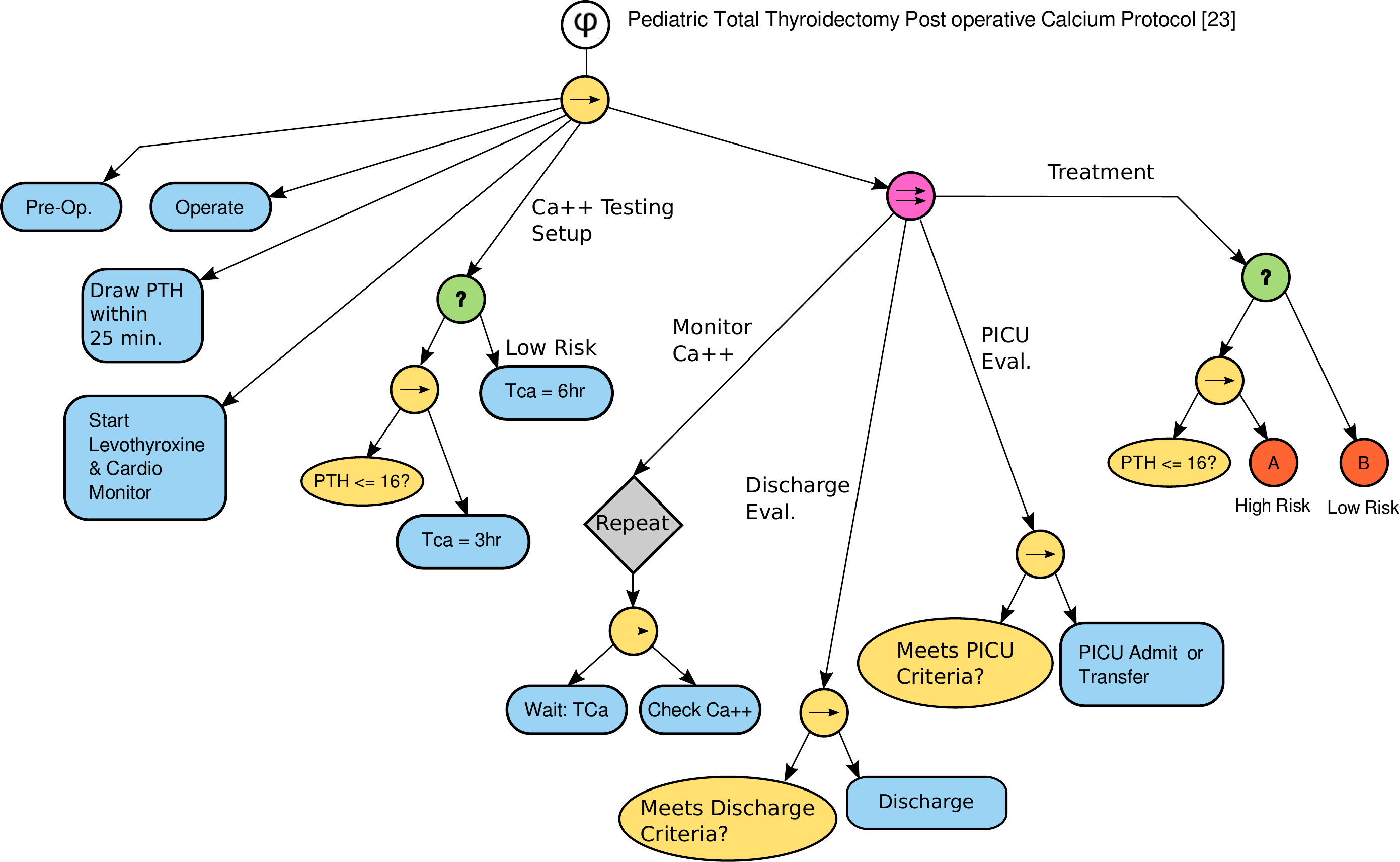}
\caption{Top Level BT for post thyroidectomy patient management of Calcium, coordinating the four BTs of Figure \ref{CaBT}.}\label{CaBT_TopLevel}
\end{figure}

Following the flowchart of Patel et al.,\cite{patel2018clinical} (Figure \ref{PatelFlowchart}),
we created a BT for this procedure by breaking it down for convenience into four smaller BTs:
\begin{enumerate}
    \item High Risk 
    \item Low Risk
    \item High Risk / Symptomatic
    \item High Risk / Asymptomatic
\end{enumerate}
Each of which has its own BT (Figure \ref{CaBT}).  
Then, completing the patient management algorithm, the four BTs of Figure \ref{CaBT}
are combined into the overall algorithm (Figure \ref{CaBT_TopLevel}).  Note that the partitioning of this BT 
as shown in Figures \ref{CaBT} and \ref{CaBT_TopLevel}
could be done in many different ways and is done here solely for graphical convenience. 

One challenging aspect of this conversion is that throughout the procedure of Figure \ref{PatelFlowchart},
there are instructions such as ``check total \Ca Q6 hours," or ``recheck total \Ca after 2 h".  Through these
distributed references,  \Ca checks are performed regularly
but are distributed throughout the flowchart and create additional process steps due to the testing time delays.  In
creating these BTs, the \Ca monitoring parts of the procedure were representationally simplified by defining

\begin{enumerate}
    \item a \Par node and BT branch to allow a repetitive \Ca monitoring process (different for High Risk and Low Risk Patients)\\
    and
    \item a variable, Tca, indicating the time between \Ca tests.
\end{enumerate}

With the above definitions, the BT turns out simpler because Tca can be modified throughout 
the procedure without otherwise storing ``state" or future
instructions for \Ca monitoring.

%%%%** Figure 7

%%%%** Figure 8

%
%              Ian not yet comfortable publishing Example 5 so delete.
%

\iffalse

\newpage
%%%%** Section 7
\section{Example 5: Management of Unilateral Antenatal Hydronephrosis}
Figure \ref{ANHProtocolFigig} illustrates a protocol in use for Unilateral Antenatal Hydronephrosis at Seattle Children's Hospital.
Part of this BT (see part D) illustrates translation of a temporal monitoring protocol from a relatively compact textual procedure,
``Renal Ultrasound every 4 months for 1 year, then every six months for 1 year, then yearly" with the use  of decorator nodes.
The of this BT section is a Repeat until success node (where the subsequent tree defines success in terms of either improvement in
SFU score, or initiating surgical treatment.   A continuation node ("F") is used to repeat the analysis and decisions for Renal Ultrasound
under each decorator.

%%%%** Figure 9
\begin{figure}[h]\centering
\includegraphics[width=0.45\textwidth]{UnilateralAntenatalHydronephrosisFLOWCHART.pdf}
\caption{}\label{ANHFlowChart}
\end{figure}

%%%%** Figure 10
\begin{figure}\centering
\includegraphics[width=0.45\textwidth]{AntenatalHydronephrosisBT_Part01.pdf}
\includegraphics[width=0.45\textwidth]{AntenatalHydronephrosisBT_Part02.pdf}
\includegraphics[width=0.45\textwidth]{AntenatalHydronephrosisBT_Part03.pdf}
\caption{BT for management of Unilateral Antenatal Hydronephrosis.}\label{ANHProtocolFigig}
\end{figure}

\fi

% \newpage
%%%%** Section 8
\section{Discussion and Conclusion}  This paper has introduced Behavior Trees as a notation and data structure for
representing medical procedures or clinical practice guidelines.  We have used some published guidelines
as examples to illustrate this representation.
The use of BTs for medical algorithms is still conceptual.  Anticipated uses to be developed
and validated in the future include:

\begin{itemize}
\item Documentation of the standard of care, through clinical practice guidelines by and for human medical providers, for example in decision support systems.
\item Integration of clinical practice guidelines with several advanced AI methods.
\item Execution frameworks for automated medical robotic tasks.
\item Description and coordination of
Human-Robot-Collaborative Systems\cite{kragic2005human,paxton2017costar}.
\end{itemize}

Compared to Finite State Machines, Hidden Markov Models, and similar
approaches, BTs afford a human-readable and writable representation through
its small number of relatively easy to understand combinatorial operators: \Seq, \Sel, and \Par,
and the ease by which BTs can be combined (using those same operators).
These properties seem to be well matched to
conventional human thinking about procedures.

There are also limitations of BTs which need further exploration and
elucidation to make sure they are used appropriately.  For example
\begin{itemize}
\item BTs do not have an explicit ``interrupt'' mechanism by which an
ongoing procedure can be stopped. For example, although the emergency airway BT of Figure \ref{airwayBTfig} illustrates
checking in parallel for the condition of low SpO2 and converting to surgical airway, if this parallel node
succeeds (i.e. the low SpO2 condition is met, and subsequent steps succeed), it does not
explicitly {\it halt} its other child tree (the remainder of the procedure on the right branch).
A similar issue exists in Figure \ref{CaBT_TopLevel} if the patient meets discharge or ICU referral criteria.

\item Human clinicians might need training if BTs are used in clincian-facing applications.
Although understanding the BT
is not difficult, and some flowcharts can be confusing, flowcharts at least have familiarity and an appearance of being self-explanatory.

\item New safety checking mechanisms (such as the ``Recovery'' node described in
Hu et al.\cite{hu2015semi,hu2017semi})  need further development and
unification.  Several authors have introduced special nodes to apply BTs in particular domains and this may require more
standardization and study.
\end{itemize}

There are exciting future possibilities for the integration of BTs with AI.  For example,
automated learning of BTs is still very much an open problem.   Initial study\cite{lim2010evolving}
and more recent works\cite{colledanchise2015learning,arXivHannafordHZL16} suggest
some possibilities for on-line autonomous performance improvement by BT driven systems.

To aid inference based on BTs, an exciting future direction is the augmentation of BTs with additional information such as probability distributions of expected duration in each leaf, or sensor emissions expected while each leaf is executes.
Such augmentation is the subject of future work.

Current CIG formalisms have been developed by academic groups to encode medical guidelines in a computer interpretable and executable manner, often to interface with medical information systems.  As such, they have depended upon research funding without clear business models\cite{SharingDream2010}.  In contrast, development of BTs has been driven by the multi-billion dollar computer gaming industry.  The demand for more realistic modeling of non-player character behavior in videogames, or recent adoption of BTs in robotics,
may spur development of additional capabilities for medical BT models.

\clearpage
\newpage
%  Use name of bibliography files without .bib extension
\bibliography{adapt_BTs_icra,Proposal,Master}
\end{document}